\documentclass[oldversion]{aa} 
\usepackage{graphicx}
\usepackage{txfonts}

\def\etal{{et al. \rm}}

\def\bcma{\object{$\beta$ CMa}}
\def\xcma{\object{$\xi^1$ CMa}}
\def\neri{\object{$\nu$ Eri}}
\def\dcet{\object{$\delta$ Cet}}
\def\voph{\object{V2052 Oph}}
\def\vcen{\object{V836 Cen}}
\def\twelvelac{\object{12 Lac}}
\def\gpeg{\object{$\gamma$ Peg}}
\def\bcep{\object{$\beta$ Cep}}
\def\gori{\object{$\gamma$ Ori}}
\def\ecma{\object{$\epsilon$ CMa}}
\def\piori{\object{$\pi^4$ Ori}}
\def\15cma{\object{15 CMa}}
\def\icma{\object{$\iota$ CMa}}
\def\tsco{\object{$\tau$ Sco}}
\def\tsco{\object{$\tau$ Sco}}

\begin{document}
   \title{
The neon content of nearby B-type stars and its implications for the solar model problem\thanks{Table~\ref{tab_ews} is only available in electronic form at the CDS via anonymous ftp to {\tt cdsarc.u-strasbg.fr (130.79.128.5)} or via {\tt http://cdsweb.u-strasbg.fr/cgi-bin/qcat?J/A+A/???/???}}}
   \titlerunning{The neon content of nearby B-type stars}

   \author{T. Morel
          \inst{1,2}
          \and
          K. Butler
          \inst{3}
          }

   \offprints{Thierry Morel, \email{morel@astro.ulg.ac.be}.}

   \institute{Institut d'Astrophysique et de G\'eophysique, Universit\'e de Li\`ege, All\'ee du 6 Ao\^ut, B\^at. B5c, 4000 Li\`ege, Belgium
         \and
         Katholieke Universiteit Leuven, Departement Natuurkunde en Sterrenkunde, Instituut voor Sterrenkunde, Celestijnenlaan 200D, B-3001 Leuven, Belgium
         \and
         Universit\"{a}ts-Sternwarte M\"{u}nchen, Scheinerstrasse 1, D-81679 M\"{u}nchen, Germany}

   \date{Received 8 April 2008 / Accepted 28 May 2008}

\abstract{The recent downward revision of the solar photospheric abundances now leads to severe inconsistencies between the theoretical predictions for the internal structure of the Sun and the results of helioseismology. There have been claims that the solar neon abundance may be underestimated and that an increase in this poorly-known quantity could alleviate (or even completely solve) this problem. Early-type stars in the solar neighbourhood are well-suited to testing this hypothesis because they are the only stellar objects whose absolute neon abundance can be derived from the direct analysis of photospheric lines. Here we present a fully homogeneous NLTE abundance study of the optical \ion{Ne}{i} and \ion{Ne}{ii} lines in a sample of 18 nearby, early B-type stars, which suggests $\log \epsilon$(Ne)=7.97$\pm$0.07 dex (on the scale in which $\log \epsilon$[H]=12) for the present-day neon abundance of the local interstellar medium (ISM). Chemical evolution models of the Galaxy only predict a very small enrichment of the nearby interstellar gas in neon over the past 4.6 Gyr, implying that our estimate should be representative of the Sun at birth. Although higher by about 35\% than the new recommended solar abundance, such a value appears insufficient by itself to restore the past agreement between the solar models and the helioseismological constraints.} 


   \maketitle
%

\section{Introduction} \label{sect_intro}
State-of-the-art spectral analyses of solar photospheric lines using time-dependent, 3-D hydrodynamical models (Asplund \etal \cite{asplund}, and references therein; hereafter AGS05) have recently led to a reduction of the commonly accepted abundances of the dominant metals in the Sun (Grevesse \& Sauval \cite{grevesse_sauval}; hereafter GS98). This, in turn, greatly affects the input physics of the standard solar models (e.g. radiative opacities) and 
considerably worsens the agreement between the theoretical predictions and
the results of helioseismic inversions. In particular, the sound speed and
density profiles in the solar interior are no longer well reproduced, while
the convective zone is predicted to be too shallow and with a helium abundance that is too low (see, e.g. Basu \& Antia \cite{basu_antia} for a comprehensive review and an account of the various solutions proposed to solve this problem). 

Neon is one of the most important contributors to the opacity at the base of the convective zone, after oxygen and iron. Contrary to these latter two elements whose abundance can be estimated from the analysis of photospheric lines (or is even accurately known from meteoritic data in the case of Fe), the Ne abundance is not well constrained. As noble gases are not retained in CI chondrite meteorites and \ion{Ne}{i} lines are completely lacking in the solar spectrum because of their high excitation energies, one has to rely instead on indirect estimates based on observations of coronal lines or high-energy particles, which are by themselves prone to large uncertainties (see below). The Ne abundance of the Sun is usually based on measurements of the [Ne/O] abundance ratio in the solar upper atmosphere and has been scaled down (by 0.17 dex) to account for the decrease in the solar oxygen content (an additional correction amounting to --0.07 dex arises from the adoption of a different neon-to-oxygen abundance ratio, as determined from energetic particles; Reames \cite{reames}). This leads to a value lowered from 8.08 (GS98) to 7.84 dex (AGS05). An upward revision of this uncertain quantity has therefore been invoked as a possible way to compensate for the decrease in opacity brought about by the lower abundances of the other chemical elements. Standard, full solar models constructed with different chemical mixtures suggest that an increase of the Ne abundance by 0.4--0.5 dex, along with a possible adjustment of the other metal abundances within their uncertainties, is required (Bahcall \etal \cite{bahcall}). Models with values outside this range do not simultaneously reproduce the helioseismic constraints that are the He abundance/depth of the convective zone and the sound-speed/density profiles in the interior (see also Delahaye \& Pinsonneault \cite{delahaye_pinsonneault}). It was also shown that an increase of this magnitude provides a better match to the properties of the solar core, as probed by low-degree {\it p}-modes (Basu \etal \cite{basu}; Zaatri \etal \cite{zaatri}). Observations of a large sample of active stars with the {\em Chandra} X-ray observatory indeed seemed at first sight to support an upward revision of the newly adopted solar Ne abundance at the required levels (Drake \& Testa \cite{drake_testa}), but much lower values were subsequently inferred for solar-like stars (e.g. Liefke \& Schmitt \cite{liefke_schmitt}) or quiescent solar regions (e.g. Young \cite{young}), thus leaving this question still open. 

The abundance patterns observed in stellar coronae appear at odds with the solar mixture, in particular, with neon being strongly enhanced in active stars by some still poorly-understood mechanisms (e.g. Drake \etal \cite{drake01}). An attractive alternative to constrain the neon content of the Sun is, however, offered by the direct analysis of \ion{Ne}{i} and \ion{Ne}{ii} photospheric lines in nearby OB stars. Although neon has traditionally been largely neglected in past abundance studies of massive stars, this 'solar model crisis' has indeed renewed interest in determining the abundance of this element in nearby objects. Since the publication of AGS05 results, however, only a single study addressing this issue appeared in the literature (Cunha \etal \cite{cunha}). A relatively high mean Ne abundance was found in a sample of 11 B-type dwarfs in the Orion association (0.27 dex above AGS05 value), but this still falls short (by a factor $\sim$1.5) of completely solving the controversy discussed above. To shed more light on this issue, here we present a fully homogeneous NLTE abundance analysis of a sample of 18 early B-type stars in the solar neighbourhood. The results presented in this paper supersede previous preliminary reports (Morel \& Butler \cite{morel_butler07}, \cite{morel_butler08}), as significant improvements in the model atom have been made during this interval.

\section{Program stars and observational material} \label{sect_obs}
The list of the program stars is provided in Table~\ref{tab_observations},
along with the source of the high-resolution spectroscopic data used.
Independent pieces of evidence have been presented for a Galactic abundance
gradient for Ne, either from observations of B stars (Kilian-Montenbruck
\etal \cite{kilian_montenbruck}), planetary nebulae (e.g. Henry \etal
\cite{henry}) or \ion{H}{ii} regions (e.g. Mart\'{\i}n-Hern\'andez \etal \cite{martin_hernandez}). Therefore, we only selected stars within $\sim$1 kpc, as estimated from {\em Hipparcos} parallaxes or open cluster membership. These stars are also uniformly distributed in terms of Galactic longitude, ensuring that the abundances we derive truly reflect those of the local ISM. The vast majority of our targets have effective temperatures lying in the range 21\,000--28\,000 K. Based on our own calculations, the \ion{Ne}{i} and \ion{Ne}{ii} lines reach a maximum strength around $T_{\rm eff}$$\sim$17\,000 and $\sim$31\,000 K, respectively. Although the diagnostic lines will be weak, our targets therefore sample a crucial $T_{\rm eff}$ range where lines of both ions are measurable and can be independently used to determine the neon abundance. As will be shown below, forcing agreement between the mean abundances provided by both ions allows one to correct for the existence of systematic errors in the data (e.g. imperfect temperature scale). 

\begin{table*}
\centering
\caption{Program stars and spectroscopic data used.}
\label{tab_observations}
\begin{tabular}{rcccccc} \hline\hline
\multicolumn{1}{c}{HD}      & \multicolumn{1}{c}{Alternative}   & Remarks$^a$ & Telescope & Instrument & Dates of         & Number of\\
\multicolumn{1}{c}{Number}  & \multicolumn{1}{c}{ID}    &             &           &                    & Observation & Exposures\\
\hline 
      886 & \gpeg       & SB1, $\beta$ Cephei & 1.9-m OHP       & ELODIE  & 1998--2004    & 47\\
    16582 & \dcet       & $\beta$ Cephei      & 1.2-m ESO/Euler & CORALIE & July 2002     & 4\\
    29248 & \neri       & $\beta$ Cephei      & 1.2-m ESO/Euler & CORALIE & 2001--2002    & 579\\
    30836 & \piori      & SB1                 & 1.2-m ESO/Euler & CORALIE & March 2007    & 4\\
    35468 & \gori       & ...                 & 1.9-m OHP       & ELODIE  & January 2003  & 2\\
    36591 & ...         & ...                 & 1.9-m OHP       & ELODIE  & November 2004 & 1\\ 
    44743 & \bcma       & $\beta$ Cephei      & 1.2-m ESO/Euler & CORALIE & 2000--2004    & 449\\ 
    46328 & \xcma       & $\beta$ Cephei      & 1.2-m ESO/Euler & CORALIE & 2000--2004    & 79\\ 
    50707 & \15cma      & $\beta$ Cephei      & 1.2-m ESO/Euler & CORALIE & March 2007    & 3\\ 
    51309 & \icma       & ...                 & 2.2-m ESO/MPG   & FEROS   & April 2005    & 6\\ 
    52089 & \ecma       & ...                 & 2.2-m ESO/MPG   & FEROS   & April 2005    & 4\\ 
 129\,929 & \vcen       & $\beta$ Cephei      & 2.2-m ESO/MPG   & FEROS   & May 2003      & 1\\ 
 149\,438 & \tsco       & ...                 & 1.2-m ESO/Euler & CORALIE & March 2007    & 2\\ 
 163\,472 & \voph       & $\beta$ Cephei      & 2.7-m McDonald  & CS2     & July 2004     & 105\\ 
 170\,580 & ...         & ...                 & 1.5-m ESO       & FEROS   & July 2001     & 1\\ 
 180\,642 & ...         & $\beta$ Cephei      & 2.2-m ESO/MPG   & FEROS   & May 2006      & 11\\ 
 205\,021 & \bcep       & SB1, $\beta$ Cephei & 1.9-m OHP       & ELODIE  & 1995--2003    & 28\\ 
 214\,993 & \twelvelac  & $\beta$ Cephei      & 2.7-m McDonald  & CS2     & July 2004     & 31\\  
\hline
\end{tabular}
\begin{flushleft}
Notes: The ELODIE spectra have been extracted from the instrument archives (see {\tt http://atlas.obs-hp.fr/elodie/}). The mean resolving power of the spectrographs amounts to about $R$$\sim$50\,000, except in the case of CS2 ($R$$\sim$60\,000). The signal-to-noise ratio in the {\em individual} exposures typically lies in the range 150--200. Further details on these observations can be found in Morel \etal (\cite{morel06}, \cite{morel08}), Morel \& Aerts (\cite{morel_aerts}), Hubrig \etal (\cite{hubrig}), and references therein.\\
$^a$ We only consider as $\beta$ Cephei stars the confirmed candidates in the recent catalogue of Stankov \& Handler (\cite{stankov_handler}).\\
\end{flushleft}
\end{table*}

Our sample comprises several pulsating B-type stars of the $\beta$ Cephei class (Table~\ref{tab_observations}). In many cases, these objects have been the targets of intensive spectroscopic monitoring campaigns dedicated to a study of their pulsational behaviour (e.g. Aerts \etal \cite{aerts04}; Mazumdar \etal \cite{mazumdar}). We make use of these data here and base our analysis in these cases on a mean spectrum created by averaging the whole set of individual exposures (all spectra were put in the laboratory rest frame prior to this operation). Using such a large (up to 579) number of time-resolved spectra not only ensures that the derived atmospheric parameters are representative of the values averaged over the pulsation cycle, but also allows us to measure with reasonable confidence some extremely weak lines with equivalent widths (EWs) down to only a few m\AA \ (see Table~\ref{tab_ews}). The accuracy of these measurements is discussed in Sect.~\ref{sect_analysis}.

Three stars in our sample are known single-lined binaries (Table~\ref{tab_observations}). The companions to \gpeg \ and \bcep \ are both very faint and the contamination of the primary spectrum can be neglected (Roberts \etal \cite{roberts}; Schnerr \etal \cite{schnerr}). In the case of \piori \ (Luyten \cite{luyten}), the lack of information about the secondary forced us to treat it as a single star (Gies \& Lambert \cite{gies_lambert} classify this system as B2 III + B2 IV, but no details are given). 

\begin{figure}
\centering
\includegraphics[width=8.0cm]{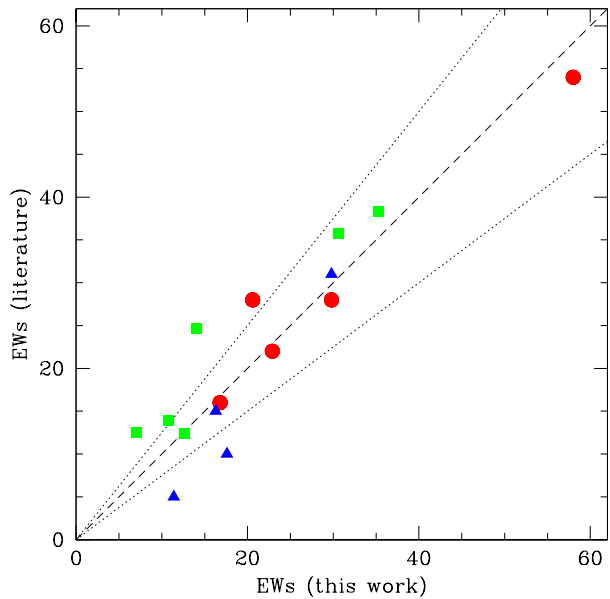}
\caption{Comparison between our measured EWs and values in the literature. {\em Circles}: Aller \& Jugaku (\cite{aller_jugaku}), but excluding uncertain values, {\em triangles}: Gies \& Lambert (\cite{gies_lambert}), {\em squares}: Kilian \& Nissen (\cite{kilian_nissen}). The 1:1 relationship is shown as a dashed line, while the two dotted lines indicate a discrepancy reaching 25\%.}
\label{fig_comp_ew}
\end{figure}

\section{Methods of analysis} \label{sect_analysis}
A standard, iterative scheme is used to self-consistently derive the atmospheric parameters purely on spectroscopic grounds: the effective temperature ($T_{\rm eff}$) is determined from the \ion{Si}{ii/iii/iv} ionization balance, the logarithmic surface gravity ($\log g$) from fitting the collisionally-broadened wings of the Balmer lines and the microturbulent velocity ($\xi$) from the \ion{O}{ii} features requiring that the abundances yielded are independent of line strength. It must be emphasised that our analysis is fully consistent, as the methodology used to derive the atmospheric parameters and elemental abundances is identical in all respects for all stars. 

Our targets are all narrow lined ($v \sin i \lesssim 65$ km s$^{-1}$) and
classical curve-of-growth techniques were used to derive the neon abundances
using the EWs of a set of \ion{Ne}{i/ii} features selected as being
unblended in the relevant temperature range. The diagnostic lines were chosen after careful inspection of spectral atlases based on the extensive line list of Kurucz \& Bell (\cite{kurucz_bell})\footnote{See the online spectral atlases of main-sequence B stars at: {\tt http://www.lsw.uni-heidelberg.de/cgi-bin/websynspec.cgi}.} and the EWs were measured by direct integration. Our final line list is made up of 12 \ion{Ne}{i} and 8 \ion{Ne}{ii} lines, and our results are based on 2 to 18 lines for a given star (see Table~\ref{tab_ews}). To roughly assess the accuracy of our EWs, we have selected a few lines of various strengths and repeated the measurements for the 79 individual exposures obtained for \xcma. The dispersion of the results was of the order of 30, 15 and 10\% for features with EW=5, 10 and 20 m\AA, respectively. This may be taken as the typical accuracy for stars with only a few spectra available, while significantly lower values are likely for stars with a very high-quality mean spectrum (see Table~\ref{tab_observations}). A comparison with the few measurements in the literature independently supports these order-of-magnitude estimates: a disagreement reaching on average about 35 and 10\% is observed for lines with EWs lower or greater than 20 m\AA, respectively (Fig.\ref{fig_comp_ew}).

To obtain the NLTE abundances, we made use of the latest versions of the line formation codes DETAIL/SURFACE and plane-parallel, fully line-blanketed, LTE atmospheric models (Kurucz \cite{kurucz93}). We adopted models with He/H=0.089 by number in all cases. The only exception was \voph, as this star displays some evidence for a helium enrichment (Morel \etal \cite{morel06}; Neiner \etal \cite{neiner}). However, adopting a solar helium abundance in that case would have a negligible impact on the resulting Ne abundance: $\Delta\log \epsilon$(Ne) $\lesssim$ 0.01 dex. 

We have developed an extensive model atom consisting of 153, 78 and 5 levels for \ion{Ne}{i}, \ion{Ne}{ii} and \ion{Ne}{iii}, respectively (along with the ground state of \ion{Ne}{iv}).\footnote{The model atoms used are available from the authors on request.}

The energy levels are those of Saloman \& Sansonetti (\cite{saloman}) for
\ion{Ne}{i}, Kramida \& Nave (\cite{kramida_a}) and Kramida \etal
(\cite{kramida_b}) for \ion{Ne}{ii}, and Kramida \& Nave (\cite{kramida_c})
for \ion{Ne}{iii}, as tabulated on the NIST web site (Ralchenko \etal \cite{ralchenko}).
The fine structure data of \ion{Ne}{ii} and \ion{Ne}{iii} have been combined
according to the LS-coupling scheme, while fine structure has been
retained for \ion{Ne}{i}.  Oscillator strengths and photoionization cross
sections for \ion{Ne}{i} and \ion{Ne}{ii} are the result of Breit-Pauli R-matrix
(BPRM; Hummer \etal \cite{hummer}) calculations (Butler \cite{butler_a},b), while those for
\ion{Ne}{iii} are from an LS-coupling R-matrix computation (Butler \cite{butler_c}).

A further BPRM calculation (Butler \cite{butler_d}) has provided collisional data for
the levels up to 5$f$ of \ion{Ne}{i}.  Griffin \etal (\cite{griffin}) performed a 61
term, 138 level, Intermediate-Coupling Frame-Transformation (ICFT) R-matrix
calculation  for \ion{Ne}{ii}.  These data provide accurate collisional 
rates for the majority of the \ion{Ne}{ii} terms considered.  All remaining
bound-bound collisional rates were estimated, either using the 
van Regemorter (\cite{vanreg}) approximation for the allowed transitions or assuming a
constant effective collision strength ($\Upsilon$=1) for the forbidden ones.
All collisional ionization rates made use of the Seaton (\cite{seaton62}) approximation
in which the photoionization cross sections at threshold were taken from the
respective R-matrix results.

Since intermediate oscillator strengths are available, we have used these
directly in the spectral synthesis.  The damping constants needed were
obtained from the R-matrix data (radiative) and from the prescription of
Cowley (\cite{cowley}; collisional).   

Although we recall that our abundance results solely rely on EW measurements and are not based on line-profile fitting techniques, the observed (weak) Ne features are well reproduced within the noise limits in all but a few cases (see Figs.\ref{fig_ne_profiles1}--\ref{fig_ne_profiles4}).

\section{Abundance results}\label{sect_results}
\begin{figure}
\centering
\includegraphics[width=9.0cm]{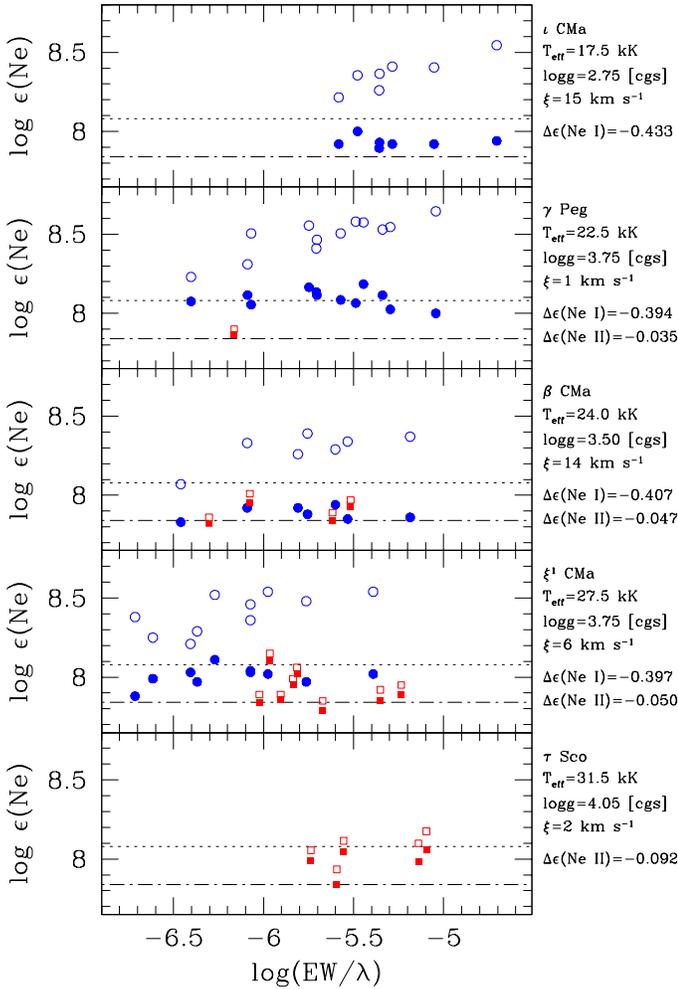}
\caption{Variations of the LTE ({\it open symbols}) and NLTE ({\it filled symbols}) abundances, as a function of the reduced EWs. The \ion{Ne}{i} and \ion{Ne}{ii} lines are denoted by {\it circles} and {\it squares}, respectively (in the online version of this journal, these are plotted in blue and red, respectively). The {\em dotted} and {\em dashed-dotted lines} indicate the solar Ne abundances of GS98 and AGS05, respectively. Each panel corresponds to a different star (ordered from top to bottom by increasing $T_{\rm eff}$). The adopted physical parameters and mean NLTE corrections for the \ion{Ne}{i} and \ion{Ne}{ii} lines, $\Delta \epsilon=\log(\epsilon)_{\rm NLTE}-\log(\epsilon)_{\rm LTE}$ (in dex), are given on the right-hand side of the panels.}
\label{fig_nlte_corrections}
\end{figure}

The \ion{Ne}{i} lines in B-type stars are known to suffer considerable NLTE strengthening (as first demonstrated by Auer \& Mihalas \cite{auer_mihalas}), leading to spuriously high LTE abundances (e.g. Gies \& Lambert \cite{gies_lambert}; Martin \cite{martin}). The NLTE corrections are much less severe in the case of the \ion{Ne}{ii} lines, however, with values typically amounting to only about --0.05 dex in our program stars, compared to --0.40 dex for the \ion{Ne}{i} lines. No strong star-to-star variations in the magnitude of these corrections are seen within our sample, either for \ion{Ne}{i} or \ion{Ne}{ii} (see Fig.\ref{fig_nlte_corrections}). The departures from LTE for the \ion{Ne}{i} lines are found to increase with the line strength (note that the strong trend between the reduced EWs and the abundances is removed under the assumption of NLTE) and reach a maximum for \ion{Ne}{i} $\lambda$6402. As Cunha \etal (\cite{cunha}), we find a small decrease of the NLTE corrections with increasing temperature for this transition, but our values appear to be systematically slightly larger (by $\sim$0.1 dex).

\begin{figure*}
\centering
\includegraphics[width=15.5cm]{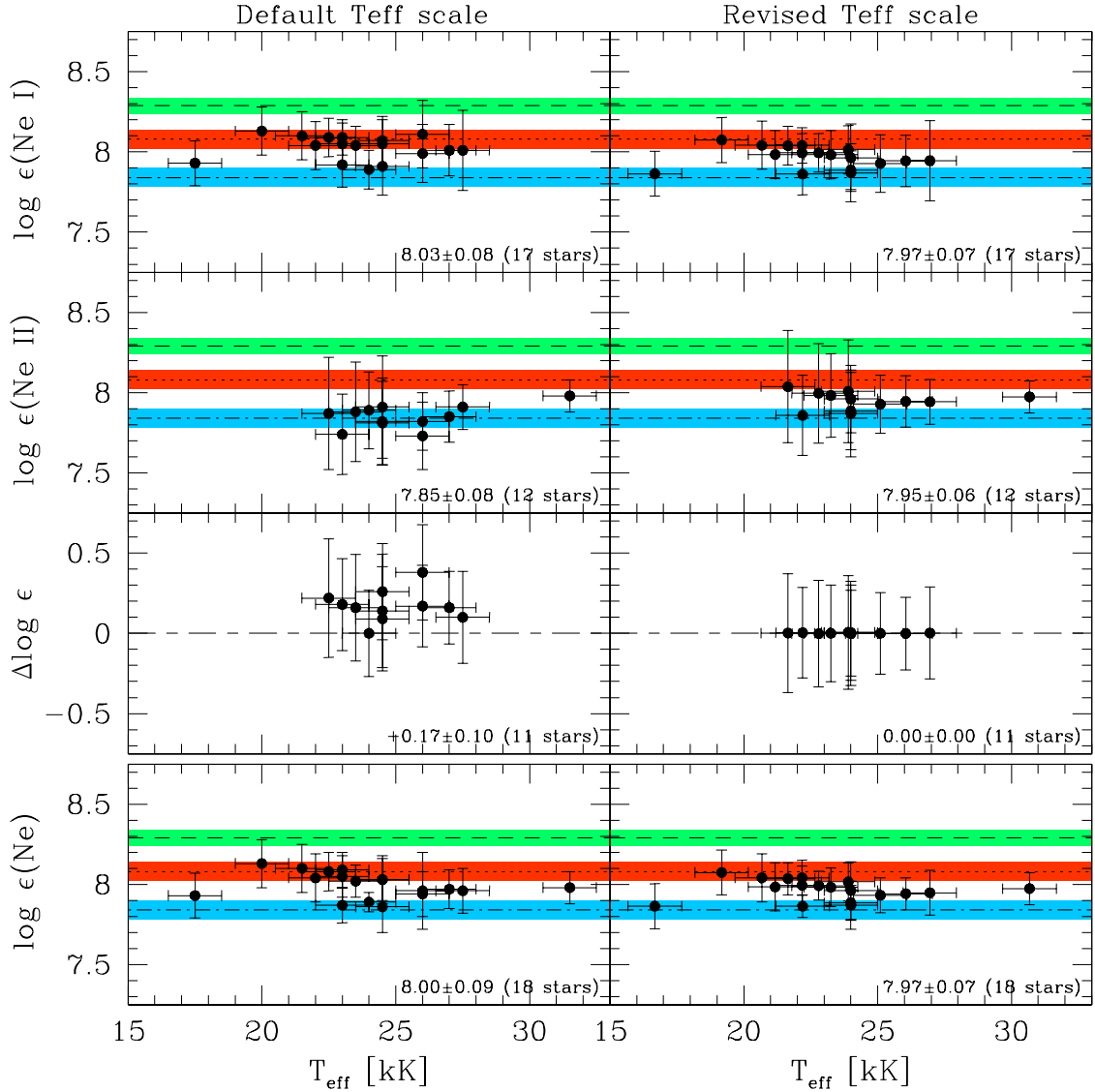}
\caption{Abundance data as a function of the effective temperature for the default ({\em left panels}) and the revised ({\em right panels}) temperature scale (see Sect.~\ref{sect_results}). From top to bottom: variation of the \ion{Ne}{i} abundances, the \ion{Ne}{ii} abundances, the difference between the abundances yielded by the \ion{Ne}{i} and \ion{Ne}{ii} lines, and the mean neon abundance using lines of both ions. The unweighted mean data value is given in each panel. The {\em long-dashed line} indicates the neon abundance needed to reconcile the solar interior models and the results of helioseismology (8.29$\pm$0.05 dex; Bahcall \etal \cite{bahcall}), while the {\em dotted} and {\em dashed-dotted lines} show the standard (8.08$\pm$0.06 dex; GS98) and the new (7.84$\pm$0.06 dex; AGS05) solar Ne abundances, respectively. The horizontal, shaded strips show the corresponding uncertainties. In the online version of this journal, these are plotted in green, red and blue, respectively.
}
\label{fig_teff}
\end{figure*}

Using the $T_{\rm eff}$ values derived from the silicon ionization balance, we find a slight, albeit clearly significant difference (0.17 dex) between the mean \ion{Ne}{i} and \ion{Ne}{ii} abundances (Fig.\ref{fig_teff}, {\it left-hand panels}). We mainly attribute this discrepancy to uncertainties in the temperature scale and not to inaccuracies in the NLTE corrections or line formation treatment. To remedy this problem, we therefore re-determined the $T_{\rm eff}$ values for the 11 stars with abundances estimated both from \ion{Ne}{i} and \ion{Ne}{ii} by imposing Ne ionization balance. This leads to a slightly cooler temperature scale with an average downward revision typical of the uncertainties and amounting to 825 K ($\sim$3.5\%). This mean value was subsequently used to rescale the effective temperatures for the 7 remaining stars with only \ion{Ne}{i} or \ion{Ne}{ii} lines measured. As the determinations of $T_{\rm eff}$ and $\log g$ are coupled, and to obtain similar quality fits to the wings of the Balmer lines, we also lowered the $\log g$ values accordingly (the microturbulent velocity was kept unchanged). The default and revised atmospheric parameters are given in Table~\ref{tab_parameters}, while the corresponding abundances are provided in Table~\ref{tab_abundances}. The quoted uncertainties take into account both the line-to-line scatter\footnote{In case where the \ion{Ne}{ii} abundance is based on a single line, we adopted a typical value of 0.10 dex.} and the errors arising from the uncertainties on the atmospheric parameters (see Morel \etal \cite{morel06}). Table~\ref{tab_errors} illustrates these calculations in the case of \bcma. The uncertainties affecting the \ion{Ne}{ii} abundances are quite large for the coolest objects in which lines of this ion could be measured (e.g. \gpeg; Table~\ref{tab_abundances}) because of the strong sensitivity of the results to errors on $T_{\rm eff}$. We regard the abundances obtained using the revised effective temperatures as the most reliable and will only discuss these values in the following, but we emphasize that the choice of the $T_{\rm eff}$ scale does not affect the conclusions drawn in this paper. As can be seen in Fig.\ref{fig_teff}, the mean Ne abundances derived using all lines only differ by 0.03 dex. Although our sample spans a wide range in terms of temperature ($\Delta$$T_{\rm eff}$$\sim$14\,000 K), no trend between the abundance data (either \ion{Ne}{i} or \ion{Ne}{ii}) and $T_{\rm eff}$ is apparent (Fig.\ref{fig_teff}, {\it right-hand panels}). The small scatter observed (standard deviation of 0.07 dex) can be explained by the measurements uncertainties alone and indicates a well-mixed local ISM with a homogeneous Ne abundance. 

\begin{table*}
\centering
\caption{Default and adopted atmospheric parameters with $T_{\rm eff}$ estimated from the Si and Ne ionization balance, respectively (see Sect.~\ref{sect_results}).}
\label{tab_parameters}
\begin{tabular}{rc|cccc|cccc} \hline\hline
\multicolumn{2}{c}{} & \multicolumn{4}{c}{Default $T_{\rm eff}$ Scale}   & \multicolumn{4}{c}{Revised $T_{\rm eff}$ Scale}\\
\multicolumn{1}{c}{HD}      & \multicolumn{1}{c}{Alternative}   & \multicolumn{1}{c}{$T_{\rm eff}$} & \multicolumn{1}{c}{$\log g$}      & \multicolumn{1}{c}{$\xi$}         & \multicolumn{1}{c}{Ref.} & \multicolumn{1}{c}{$T_{\rm eff}$} & \multicolumn{1}{c}{$\log g$}      & \multicolumn{1}{c}{$\xi$}         & \multicolumn{1}{c}{Ref.}\\
\multicolumn{1}{c}{Number}  & ID    & \multicolumn{1}{c}{(K)}           & \multicolumn{1}{c}{(cm s$^{-2}$)} & \multicolumn{1}{c}{(km s$^{-1}$)} & \multicolumn{1}{c}{} & \multicolumn{1}{c}{(K)}           & \multicolumn{1}{c}{(cm s$^{-2}$)} & \multicolumn{1}{c}{(km s$^{-1}$)} & \multicolumn{1}{c}{}   \\
\hline 
      886 & \gpeg       & 22\,500 & 3.75$\pm$0.15 & 1$^{+2}_{-1}$ & 1 & 21\,650 & 3.67$\pm$0.15 & 1$^{+2}_{-1}$ & 5\\	 
    16582 & \dcet       & 23\,000 & 3.80$\pm$0.15 & 1$^{+3}_{-1}$ & 1 & 22\,175 & 3.72$\pm$0.15 & 1$^{+3}_{-1}$ & 5\\	 
    29248 & \neri       & 23\,500 & 3.75$\pm$0.15 & 10$\pm$4      & 1 & 22\,800 & 3.68$\pm$0.15 & 10$\pm$4      & 5\\	 
    30836 & \piori      & 21\,500 & 3.35$\pm$0.15 &  8$\pm$4      & 2 & 20\,675 & 3.27$\pm$0.15 &  8$\pm$4      & 5\\	 
    35468 & \gori       & 22\,000 & 3.50$\pm$0.20 & 13$\pm$5      & 2 & 21\,175 & 3.42$\pm$0.20 & 13$\pm$5      & 5\\	 
    36591 & ...         & 27\,000 & 4.00$\pm$0.20 & 3$\pm$2       & 2 & 26\,050 & 3.91$\pm$0.20 & 3$\pm$2       & 5\\   
    44743 & \bcma       & 24\,000 & 3.50$\pm$0.15 & 14$\pm$3      & 1 & 24\,000 & 3.50$\pm$0.15 & 14$\pm$3      & 5\\   
    46328 & \xcma       & 27\,500 & 3.75$\pm$0.15 & 6$\pm$2       & 1 & 26\,950 & 3.70$\pm$0.15 & 6$\pm$2       & 5\\   
    50707 & \15cma      & 26\,000 & 3.60$\pm$0.15 & 7$\pm$3       & 2 & 24\,000 & 3.40$\pm$0.15 & 7$\pm$3       & 5\\   
    51309 & \icma       & 17\,500 & 2.75$\pm$0.15 & 15$\pm$5      & 2 & 16\,675 & 2.67$\pm$0.15 & 15$\pm$5      & 5\\   
    52089 & \ecma       & 23\,000 & 3.30$\pm$0.15 & 16$\pm$4      & 2 & 22\,200 & 3.22$\pm$0.15 & 16$\pm$4      & 5\\   
 129\,929 & \vcen       & 24\,500 & 3.95$\pm$0.20 & 6$\pm$3       & 1 & 23\,900 & 3.89$\pm$0.20 & 6$\pm$3       & 5\\   
 149\,438 & \tsco       & 31\,500 & 4.05$\pm$0.15 & 2$\pm$2       & 3 & 30\,675 & 3.97$\pm$0.15 & 2$\pm$2       & 5\\   
 163\,472 & \voph       & 23\,000 & 4.00$\pm$0.20 & 1$^{+4}_{-1}$ & 1 & 22\,175 & 3.92$\pm$0.20 & 1$^{+4}_{-1}$ & 5\\   
 170\,580 &  ...        & 20\,000 & 4.10$\pm$0.15 & 1$_{-1}^{+5}$ & 4 & 19\,175 & 4.02$\pm$0.15 & 1$_{-1}^{+5}$ & 5\\   
 180\,642 &  ...        & 24\,500 & 3.45$\pm$0.15 & 12$\pm$3      & 4 & 24\,000 & 3.40$\pm$0.15 & 12$\pm$3      & 5\\   
 205\,021 & \bcep       & 26\,000 & 3.70$\pm$0.15 & 6$\pm$3       & 1 & 25\,100 & 3.61$\pm$0.15 & 6$\pm$3       & 5\\   
 214\,993 & \twelvelac  & 24\,500 & 3.65$\pm$0.15 & 10$\pm$4      & 1 & 23\,250 & 3.53$\pm$0.15 & 10$\pm$4      & 5\\    
\hline
\end{tabular}
\begin{flushleft}
Notes: The 1-$\sigma$ uncertainty on $T_{\rm eff}$ is 1000 K for all stars.\\
References. (1) Morel \etal \cite{morel06}; (2) Morel \etal \cite{morel08}; (3) Hubrig \etal \cite{hubrig}; (4) Morel \& Aerts \cite{morel_aerts}; (5) this paper.\\
\end{flushleft}
\end{table*}

\begin{table*}
\centering
\caption{Derived neon abundances.}
\hspace*{-0.3cm}
\label{tab_abundances}
\begin{tabular}{rc|cccccc|cccccc} \hline\hline
\multicolumn{2}{c}{} & \multicolumn{6}{c}{Default $T_{\rm eff}$ Scale}   & \multicolumn{6}{c}{Revised $T_{\rm eff}$ Scale}\\
\multicolumn{1}{c}{HD}      & Alternative  & \multicolumn{2}{c}{$\log \epsilon$(\ion{Ne}{i})} & \multicolumn{2}{c}{$\log \epsilon$(\ion{Ne}{ii})} &
\multicolumn{2}{c}{$\log \epsilon$(Ne)}  & \multicolumn{2}{c}{$\log
\epsilon$(\ion{Ne}{i})} & \multicolumn{2}{c}{$\log \epsilon$(\ion{Ne}{ii})} &
\multicolumn{2}{c}{$\log \epsilon$(Ne)} \\
\multicolumn{1}{c}{Number}  & ID   & \multicolumn{2}{c}{(dex)}
&  \multicolumn{2}{c}{(dex)}                 & \multicolumn{2}{c}{(dex)}                &
\multicolumn{2}{c}{(dex)}                 &  \multicolumn{2}{c}{(dex)}                 &
\multicolumn{2}{c}{(dex)}\\
\hline
      886 & \gpeg       & 8.09$\pm$0.12 & (12) &  7.87$\pm$0.35 & (1) &  8.08$\pm$0.12 & (13) & 8.04$\pm$0.12 & (12) & 8.04$\pm$0.35 & (1) & 8.04$\pm$0.10 & (13)\\
    16582 & \dcet       & 8.05$\pm$0.13 & (11) &  ...           & ... &  8.05$\pm$0.13 & (11) & 7.99$\pm$0.12 & (11) & ...           & ... & 7.99$\pm$0.12 & (11)\\
    29248 & \neri       & 8.04$\pm$0.12 & (8)  &  7.88$\pm$0.31 & (1) &  8.02$\pm$0.10 & (9)  & 7.99$\pm$0.12 & (8)  & 7.99$\pm$0.31 & (1) & 7.99$\pm$0.09 & (9) \\
    30836 & \piori      & 8.10$\pm$0.15 & (7)  &  ...           & ... &  8.10$\pm$0.15 & (7)  & 8.04$\pm$0.15 & (7)  & ...           & ... & 8.04$\pm$0.15 & (7) \\
    35468 & \gori       & 8.04$\pm$0.15 & (5)  &  ...           & ... &  8.04$\pm$0.15 & (5)  & 7.98$\pm$0.15 & (5)  & ...           & ... & 7.98$\pm$0.15 & (5) \\ 
    36591 & ...         & 8.01$\pm$0.16 & (6)  &  7.85$\pm$0.16 & (2) &  7.97$\pm$0.12 & (8)  & 7.94$\pm$0.16 & (6)  & 7.94$\pm$0.16 & (2) & 7.94$\pm$0.10 & (8) \\ 
    44743 & \bcma       & 7.89$\pm$0.12 & (7)  &  7.89$\pm$0.24 & (4) &  7.89$\pm$0.06 & (11) & 7.89$\pm$0.12 & (7)  & 7.89$\pm$0.24 & (4) & 7.89$\pm$0.11 & (11)\\ 
    46328 & \xcma       & 8.01$\pm$0.25 & (10) &  7.91$\pm$0.14 & (8) &  7.96$\pm$0.14 & (18) & 7.94$\pm$0.25 & (10) & 7.94$\pm$0.14 & (8) & 7.94$\pm$0.14 & (18)\\ 
    50707 & \15cma      & 8.11$\pm$0.21 & (3)  &  7.73$\pm$0.21 & (2) &  7.96$\pm$0.24 & (5)  & 7.96$\pm$0.21 & (3)  & 7.96$\pm$0.21 & (2) & 7.96$\pm$0.18 & (5) \\ 
    51309 & \icma       & 7.93$\pm$0.14 & (7)  &  ...           & ... &  7.93$\pm$0.14 & (7)  & 7.86$\pm$0.14 & (7)  & ...           & ... & 7.86$\pm$0.14 & (7) \\
    52089 & \ecma       & 7.92$\pm$0.14 & (6)  &  7.74$\pm$0.25 & (2) &  7.87$\pm$0.11 & (8)  & 7.86$\pm$0.13 & (6)  & 7.86$\pm$0.25 & (2) & 7.86$\pm$0.07 & (8) \\ 
 129\,929 & \vcen       & 8.05$\pm$0.15 & (6)  &  7.91$\pm$0.32 & (1) &  8.03$\pm$0.13 & (7)  & 8.01$\pm$0.15 & (6)  & 8.01$\pm$0.32 & (1) & 8.01$\pm$0.12 & (7) \\ 
 149\,438 & \tsco       & ...           &  ... &  7.98$\pm$0.10 & (5) &  7.98$\pm$0.10 & (5)  & ...           &  ... & 7.97$\pm$0.10 & (5) & 7.97$\pm$0.10 & (5) \\  
 163\,472 & \voph       & 8.09$\pm$0.11 & (2)  &  ...           & ... &  8.09$\pm$0.11 & (2)  & 8.04$\pm$0.11 & (2)  & ...           & ... & 8.04$\pm$0.11 & (2) \\ 
 170\,580 & ...         & 8.13$\pm$0.15 & (11) &  ...           & ... &  8.13$\pm$0.15 & (11) & 8.07$\pm$0.14 & (11) & ...           & ... & 8.07$\pm$0.14 & (11)\\ 
 180\,642 & ...         & 7.91$\pm$0.18 & (3)  &  7.82$\pm$0.27 & (3) &  7.86$\pm$0.16 & (6)  & 7.87$\pm$0.18 & (3)  & 7.87$\pm$0.27 & (3) & 7.87$\pm$0.15 & (6) \\ 
 205\,021 & \bcep       & 7.99$\pm$0.18 & (5)  &  7.82$\pm$0.18 & (2) &  7.94$\pm$0.14 & (7)  & 7.93$\pm$0.18 & (5)  & 7.93$\pm$0.18 & (2) & 7.93$\pm$0.11 & (7) \\ 
 214\,993 & \twelvelac  & 8.07$\pm$0.15 & (6)  &  7.81$\pm$0.26 & (1) &  8.03$\pm$0.15 & (7)  & 7.98$\pm$0.15 & (6)  & 7.98$\pm$0.26 & (1) & 7.98$\pm$0.12 & (7) \\
\hline
\end{tabular}
\begin{flushleft}
Notes: The abundances are given for both the default (estimated from the Si ionization balance) and the revised $T_{\rm eff}$ scales (estimated from the Ne ionization balance for the stars with both \ion{Ne}{i} and \ion{Ne}{ii} abundances, and offset by --825 K otherwise; see Sect.~\ref{sect_results}). In each case we provide the abundances yielded by the \ion{Ne}{i} and \ion{Ne}{ii} lines alone, and the values derived using all lines (on the scale in which $\log \epsilon$[H]=12). The number of lines used for each ion is given in the parentheses. A blank indicates that the spectral features pertaining to a given ionization stage were not measurable.\\
\end{flushleft}
\end{table*}

\section{Discussion}\label{sect_discussion}
\subsection{Comparison with previous results for B-type stars}\label{sect_comparison}
A mean neon abundance, $\log \epsilon$(Ne)=7.97$\pm$0.07 dex, is inferred from our combined NLTE abundance analysis of the \ion{Ne}{i} and \ion{Ne}{ii} lines (a nearly identical result to within 0.01 dex is obtained when weighting the individual abundances by their respective uncertainties). We compare below our results with those reported in the literature from the NLTE analysis of the \ion{Ne}{i} lines. To our knowledge, only a single study based on \ion{Ne}{ii} has been performed (Kilian \cite{kilian94}). Although LTE was assumed, these results will also be discussed in the following in view of the smallness of the NLTE corrections for this ion (Sect.~\ref{sect_results}).

First, a significantly higher value, $\log \epsilon$(Ne)=8.11$\pm$0.04 dex, has recently been reported from the analysis of 8 \ion{Ne}{i} transitions in 11 Orion B-type dwarfs by  Cunha \etal (\cite{cunha}) using an extensive TLUSTY model atom and NLTE model atmospheres (Hubeny \& Lanz \cite{hubeny_lanz}). The adopted $T_{\rm eff}$ and $\log g$ values were taken from Cunha \& Lambert (\cite{cunha_lambert}). The effective temperatures were estimated from Str\"omgren indices and the surface gravities from fitting the collisionally-broadened wings of H$\gamma$. One star (\object{HD 36591}) is common to both studies and follows this general trend, as we obtain a nominal value lower by 0.14 dex. This discrepancy may have various causes and can stem from: (a) the magnitude of the applied NLTE corrections: as mentioned in Sect.~\ref{sect_results}, and although this conclusion is only based on a single transition, there is some indication that ours are slightly larger; (b) differences in the analytical procedures used to derive the atmospheric parameters\footnote{We note that some of their adopted surface gravities are uncomfortably large: up to $\log g$=4.70 dex in the complete sample analysed by Cunha \& Lambert (\cite{cunha_lambert}) and up to $\log g$=4.41 dex in the sample of Cunha \etal (\cite{cunha}). The atmospheric parameters obtained for \object{HD 36591} are, however, in reasonable agreement: $T_{\rm eff}$=27\,000$\pm$1000 K and $\log g$=4.00$\pm$0.20 dex (this study) against $T_{\rm eff}$=26\,330$\pm$750 K and $\log g$=4.21$\pm$0.10 dex (Cunha \etal \cite{cunha}). Our microturbulent velocity ($\xi$=3$\pm$2 km s$^{-1}$) is very close to their fixed value, $\xi$=5 km s$^{-1}$, but the Ne abundances are, in any case, largely insensitive to this quantity (Table~\ref{tab_errors}).}; (c) the choice of the atomic data: Table~\ref{loggf} shows, however, that our $\log gf$ values are in good agreement with theirs (taken from Seaton \cite{seaton}); or (d) the type of model atmosphere employed (LTE vs NLTE). Nieva \& Przybilla (\cite{nieva_przybilla}) carried out a thorough comparison in the case of OB dwarfs and giants between the model predictions obtained assuming line-blanketed, LTE model atmospheres coupled with a detailed NLTE line-formation treatment (as in our study) or a self-consistent, full NLTE approach (as in Cunha \etal \cite{cunha}). Excellent agreement was found for \tsco \ both for the atmospheric structure (e.g. temperatures deviating by less than 1\% in the Ne line-formation regions) and for the theoretical H and He line profiles, which were found to be nearly indistinguishable when using identical model atoms. Even smaller discrepancies are expected for the other stars in our sample, as NLTE effects are very likely to be less critical in cooler dwarfs. Although detailed calculations (beyond the scope of this paper) are needed to quantitatively examine this issue, this supports the idea that the hybrid approach adopted here is suitable for an accurate determination of Ne abundances in B dwarfs at solar metallicity (and additionally much less demanding in terms of computer resources). 

\begin{table}
\centering
\caption{Calculation of the error budget in the case of \bcma \ (and with the default $T_{\rm eff}$ scale).}
\label{tab_errors}
\begin{tabular}{lrrr} \hline\hline
                              & \multicolumn{1}{c}{$\Delta\log \epsilon$(\ion{Ne}{i})} & \multicolumn{1}{c}{$\Delta\log \epsilon$(\ion{Ne}{ii})} & \multicolumn{1}{c}{$\Delta$$\log\epsilon$(Ne)}\\
                              & \multicolumn{1}{c}{(dex)} & \multicolumn{1}{c}{(dex)} & \multicolumn{1}{c}{(dex)}\\
\hline
$\sigma_{\rm int}$            &   0.042 &   0.065 &   0.048\\ 
$\sigma_{T_{\rm eff}}$        &  +0.087 & --0.175 & --0.008\\ 
$\sigma_{\log g}$             & --0.012 &  +0.067 &  +0.018\\ 
$\sigma_{\xi}$                & --0.004 & --0.012 & --0.008\\ 
$\sigma_{T_{\rm eff}/\log g}$ &  +0.068 & --0.117 &  +0.001\\ 
$\sigma_T$                    &   0.119 &   0.231 &   0.053\\
\hline
\end{tabular}
\begin{flushleft}
Notes: $\sigma_{\rm int}$: line-to-line scatter; $\sigma_{T_{\rm eff}}$: variation of the abundances for $\Delta T_{\rm eff}$=+1000 K; $\sigma_{\log g}$: as before, but for $\Delta \log g$=+0.15 dex; $\sigma_{\xi}$: as before, but for $\Delta \xi$=+3 km s$^{-1}$; $\sigma_{T_{\rm eff}/\log g}$: as before, but for $\Delta T_{\rm eff}$=+1000 K and $\Delta \log g$=+0.15 dex; $\sigma_T$: total uncertainty.\\
\end{flushleft}
\end{table}

An NLTE neon abundance study of 9 bright B5--B9 stars was also presented by Hempel \& Holweger (\cite{hempel_holweger}). After deriving $T_{\rm eff}$ and $\log g$ from photometric data, they obtained  $\log \epsilon$(Ne)=8.16$\pm$0.14 dex from the \ion{Ne}{i} $\lambda$6402 and \ion{Ne}{i} $\lambda$6506 lines using a model atom consisting of 45 and 47 levels for \ion{Ne}{i} and \ion{Ne}{ii}, respectively. 

Using a 31-level \ion{Ne}{i} model atom and TLUSTY NLTE model atmospheres, Dworetsky \& Budaj (\cite{dworetsky_budaj}) obtained $\log \epsilon$(Ne)=8.10$\pm$0.09 dex for 7 late B/early A stars based on the modelling of \ion{Ne}{i} $\lambda$6402. The atmospheric parameters were estimated from a combination of photometric and spectroscopic techniques. 

On the other hand, Sigut (\cite{sigut}) developed a model atom consisting of 37 \ion{Ne}{i} and 11 \ion{Ne}{ii} levels to recompute the NLTE abundances of 14 early B-type stars in the sample of Gies \& Lambert (\cite{gies_lambert}) using their EWs for \ion{Ne}{i} $\lambda$6506 and adopted atmospheric parameters. A mean value, $\log \epsilon$(Ne)$\sim$8.12 dex, was obtained, but the scatter is very large (spread of 0.7 dex). Furthermore, the $T_{\rm eff}$ scale assumed by Gies \& Lambert (\cite{gies_lambert}) is likely too hot (see discussion in Lyubimkov \etal \cite{lyubimkov}).  

Finally, Kilian (\cite{kilian94}) derived the Ne abundance of 12 nearby, early B-type stars using a set of \ion{Ne}{ii} lines and found a mean value: $\log \epsilon$(Ne)=8.10$\pm$0.06 dex.\footnote{We do not discuss the data for the distant open clusters \object{NGC 6231}, \object{NGC 6611}, \object{S285} and \object{S289} (Kilian \etal \cite{kilian_etal94}; Kilian-Montenbruck \etal \cite{kilian_montenbruck}) because of the very likely existence of a Galactic Ne abundance gradient (see Sect.~\ref{sect_obs}).} The atmospheric parameters were estimated from the NLTE analysis of the H and Si line profiles, but it should be noted that atmospheric models with an inadequate treatment of metal line blanketing were used (Gold \cite{gold}). Considering the similar nature of that sample and ours, we can expect the NLTE corrections affecting the \ion{Ne}{ii} transitions to be of similar magnitude and, as a consequence, to be also small. Assuming $\Delta \epsilon$$\sim$--0.05 dex (Sect.~\ref{sect_results}) would lead to a corrected abundance, $\log \epsilon$(Ne)$\sim$8.05 dex, again slightly higher than our best mean estimate. Two stars are in common with our study: \object{HD 36591} and \tsco. We find an Ne abundance lower by 0.11 and 0.18 dex, respectively. Even accounting for the small NLTE corrections discussed above, our values appear indeed slightly lower even on a star-to-star basis. 

\begin{table*}
\centering
\caption{Comparison of the $\log gf$ values used here and those of Seaton (\cite{seaton}).}
\label{loggf}
\hspace*{-0.3cm}
\begin{tabular}{rrrrr} \hline\hline
\multicolumn{1}{c}{Lower Level} & \multicolumn{1}{c}{Upper Level} & \multicolumn{1}{c}{$\lambda$} & \multicolumn{2}{c}{$\log gf$} \\
                                &                                 & \multicolumn{1}{c}{(\AA)}     & \multicolumn{1}{c}{This work} & \multicolumn{1}{c}{Seaton} \\
\hline
2p$^5$($^2$P$^{\rm o}_{3/2}$)3s[3/2]$_2$ & 2p$^5$($^2$P$^{\rm o}_{3/2}$)3p[3/2]$_2$ & 6143.063 & $-0.069$ &  $-0.101$ \\
2p$^5$($^2$P$^{\rm o}_{1/2}$)3s[1/2]$_0$ & 2p$^5$($^2$P$^{\rm o}_{1/2}$)3p[1/2]$_1$ & 6163.594 & $-0.613$ &  $-0.616$ \\
2p$^5$($^2$P$^{\rm o}_{1/2}$)3s[1/2]$_0$ & 2p$^5$($^2$P$^{\rm o}_{1/2}$)3p[3/2]$_1$ & 6266.495 & $-0.334$ &  $-0.370$ \\
2p$^5$($^2$P$^{\rm o}_{3/2}$)3s[3/2]$_2$ & 2p$^5$($^2$P$^{\rm o}_{3/2}$)3p[5/2]$_2$ & 6334.428 & $-0.324$ &  $-0.315$ \\
2p$^5$($^2$P$^{\rm o}_{3/2}$)3s[3/2]$_1$ & 2p$^5$($^2$P$^{\rm o}_{3/2}$)3p[3/2]$_1$ & 6382.991 & $-0.228$ &  $-0.236$ \\
2p$^5$($^2$P$^{\rm o}_{3/2}$)3s[3/2]$_2$ & 2p$^5$($^2$P$^{\rm o}_{3/2}$)3p[5/2]$_3$ & 6402.248 & $ 0.336$ &  $ 0.330$ \\
2p$^5$($^2$P$^{\rm o}_{3/2}$)3s[3/2]$_1$ & 2p$^5$($^2$P$^{\rm o}_{3/2}$)3p[5/2]$_2$ & 6506.528 & $-0.026$ &  $-0.033$ \\
\hline
\end{tabular}
\end{table*}

Summarizing, the discussion above shows that the previous determinations in the literature cluster around $\log \epsilon$(Ne)$\sim$8.1 dex and that our mean abundance lies at the lower end of the proposed values (although marginally compatible within the errors). However, it has to be kept in mind that our study is the first one requiring agreement between two adjacent Ne ionization stages. Had we only used the \ion{Ne}{i} lines, without any constraints imposed by the \ion{Ne}{ii} lines, we would have inferred a value, $\log \epsilon$(Ne)=8.03$\pm$0.08 dex, in better agreement with the studies discussed above (Fig.\ref{fig_teff}; {\it upper, left-hand panel}). 
 
\subsection{The Ne content of the Sun from different indicators}\label{sect_sun}
Our mean Ne abundance is intermediate between the standard solar Ne abundance (8.08$\pm$0.06 dex; GS98) and the most recent estimate following the decrease in the oxygen content (7.84$\pm$0.06 dex; AGS05), and therefore does not support a high neon abundance for the Sun (we note that in principle a more adequate comparison should be made with the Ne abundance at the base of the solar convective zone, which is thought to be some 0.04 dex higher than the surface value because of gravitational element settling; Turcotte \etal \cite{turcotte}).  

A reexamination of this quantity was largely motivated by the analysis of highly-ionized \ion{Ne}{ix/x} lines in the high-resolution spectra of 21 (mostly) strong stellar X-ray emitters collected by the {\em Chandra} satellite (Drake \& Testa \cite{drake_testa}), which suggested a [Ne/O] abundance ratio in the solar corona much higher than the value adopted by AGS05 (taken from Reames \cite{reames}): $\sim$0.41 vs $\sim$0.15. Employing such a high ratio, along with the recently revised solar oxygen abundance ($\log \epsilon$[O]$\sim$8.66 dex; AGS05), would imply $\log \epsilon$(Ne)$\sim$8.27 dex for the Sun, a value that is potentially high enough to allow the standard solar models to fulfil the helioseismological constraints (8.29$\pm$0.05 dex; Bahcall \etal \cite{bahcall}).  

Numerous works have demonstrated a dependence between the abundances of the chemical elements and their first ionization potential (FIP) in stellar transition regions and coronae, with the nature of this relationship depending, however, on the stellar activity level. This phenomenon is still not well understood and a variety of behaviours are observed, but some general trends are emerging (e.g. Audard \etal \cite{audard}; Telleschi \etal \cite{telleschi}). First, in relatively quiescent stars (such as the Sun) elements with low FIPs ($\lesssim$10 eV; e.g. Mg, Fe) tend to be overabundant compared to the photospheric reference values.  In contrast, an opposite trend is observed in strong coronal sources, where these chemical species are depleted with respect to the high-FIP elements (e.g. N, O, Ne). In view of the significantly different FIPs of O and Ne (13.6 and 21.6 eV, respectively), a concern is that these elements may suffer various levels of chemical fractionation in the active stars analysed by Drake \& Testa (\cite{drake_testa}). In moderately active coronal sources or stars in the higher tail of the stellar activity distribution (such as young, fast-rotating dwarfs or some RS CVn binaries in the sample of Drake \& Testa \cite{drake_testa}), high [Ne/O] abundance ratios are commonly observed (e.g. Brinkman \etal \cite{brinkman}; Wood \& Linsky \cite{wood_linsky}). As shown by G\"udel (\cite{guedel}), this ratio increases in parallel with the characteristic temperature of the plasma in stellar coronae (which is itself correlated with the activity level, as parametrised, for instance, by the ratio of the X-ray to bolometric luminosities). Impulsive, Ne-rich flares are also observed in the Sun, with [Ne/O] enhancements reaching up to a factor $\sim$2 (Schmelz \cite{schmelz93}). A similar situation might prevail in very active stars whose upper atmosphere is believed to be permeated by strong flaring events. Drake \& Testa (\cite{drake_testa}) addressed this problem and dismissed the possibility that it could affect their conclusions on the basis of the similar [Ne/O] ratios obtained for a few stars with low-activity levels. This issue remains controversial, however, as other studies derived much lower ratios in this activity regime (e.g. Liefke \& Schmitt \cite{liefke_schmitt}; Raassen \etal \cite{raassen02}). As illustrated by the supplementary data provided by Drake \& Testa (\cite{drake_testa}), even in the Sun the [Ne/O] measurements in the outer atmosphere show considerable scatter and may depend on the region under investigation (e.g. active regions, flares, wind). Although observations of the quiet Sun clearly favour a low ratio (e.g. Young \cite{young}), definitive statements regarding the  value of [Ne/O] in the underlying photosphere must probably await a better theoretical understanding of such behaviour. Another limitation affecting the solar Ne abundance derived from X-ray data is that this quantity is tied to the photospheric oxygen abundance, whose exact value is still being debated (e.g. Antia \& Basu \cite{antia_basu06}). Taking an element with a well-defined meteoritic abundance as reference (e.g. Mg) would avoid this problem, but the abundance ratios of these species relative to Ne are unfortunately more likely to be affected by FIP effects. Taking these two sources of uncertainty together, it might thus prove difficult in the foreseeable future to set very tight constraints on the Ne content of the Sun from coronal observations alone. 

\begin{figure}
\centering
\includegraphics[width=9.0cm]{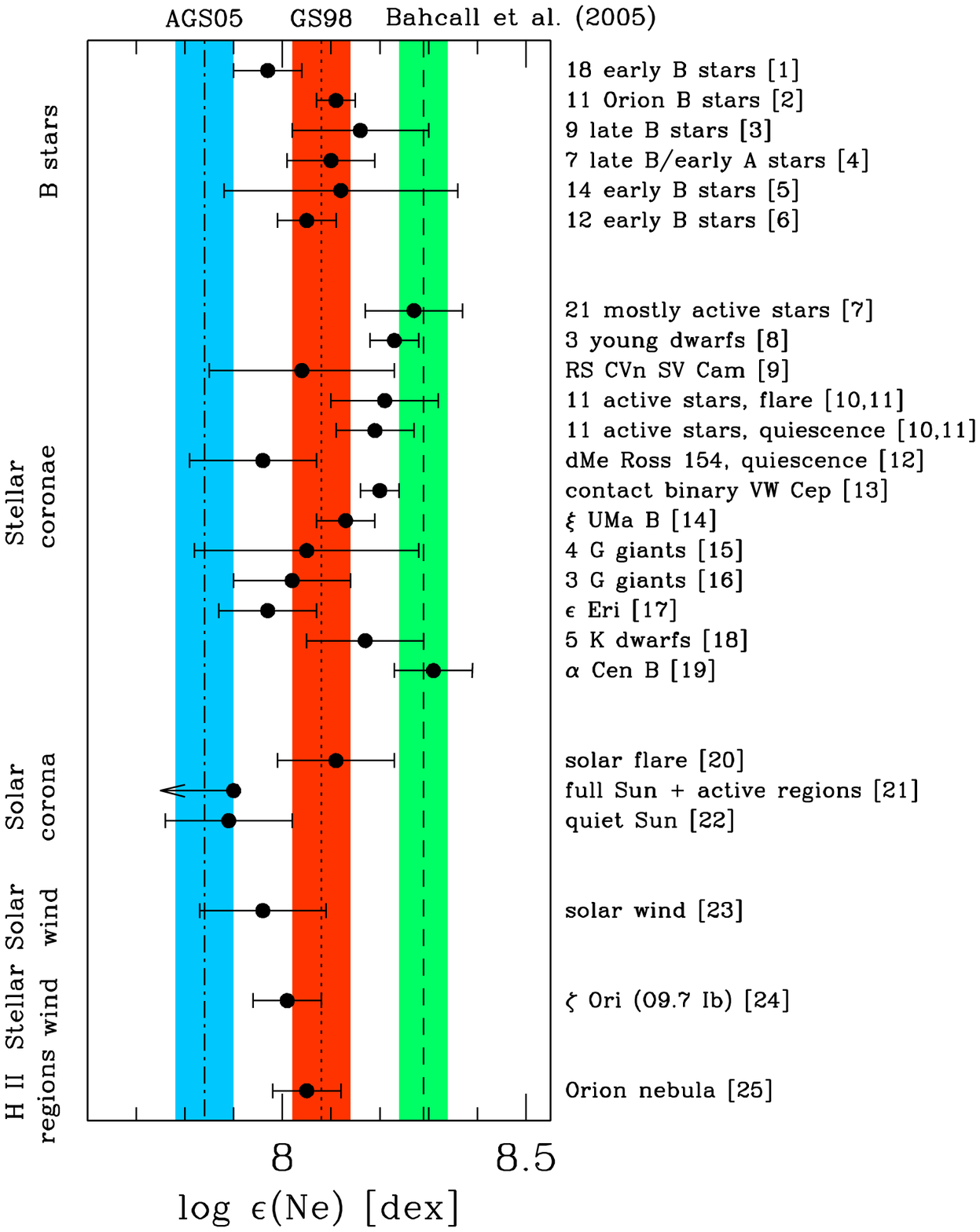}
\caption{Neon abundances from different indicators recently reported in the literature. The stellar coronal sources are roughly ordered as a function of decreasing activity from top to bottom. The {\em long-dashed line} indicates the Ne abundance needed to reconcile the solar interior models and the results of helioseismology (8.29$\pm$0.05 dex; Bahcall \etal \cite{bahcall}), while the {\em dotted} and {\em dashed-dotted lines} show the standard (8.08$\pm$0.06 dex; GS98) and the new (7.84$\pm$0.06 dex; AGS05) solar Ne abundances, respectively. The vertical, shaded strips show the corresponding uncertainties. In the online version of this journal, these are plotted in green, red and blue, respectively. 
References. (1) this paper; (2) Cunha \etal \cite{cunha}; (3) Hempel \& Holweger \cite{hempel_holweger}; (4) Dworetsky \& Budaj \cite{dworetsky_budaj}; (5) Sigut \cite{sigut}; (6) Kilian \cite{kilian94}, but corrected by --0.05 dex (see text); (7) Drake \& Testa \cite{drake_testa}; (8) Garc\'{\i}a-Alvarez \etal \cite{garcia08}; (9) Sanz-Forcada \etal \cite{sanz_forcada}; (10) Nordon \& Behar \cite{nordon_behar07}; (11) Nordon \& Behar \cite{nordon_behar08}; (12) Wargelin \etal \cite{wargelin}; (13) Huenemoerder \etal \cite{huenemoerder}; (14) Ball \etal \cite{ball}; (15) Ayres \etal \cite{ayres}; (16) Garc\'{\i}a-Alvarez \etal \cite{garcia06}; (17) Ness \& Jordan \cite{ness_jordan};  (18) Wood \& Linsky \cite{wood_linsky}; (19) Liefke \& Schmitt \cite{liefke_schmitt}; (20) Landi \etal \cite{landi}; (21) Schmelz \etal \cite{schmelz05} (upper limit); (22) Young \cite{young}; (23) Bochsler \cite{bochsler}; (24) Raassen \etal \cite{raassen08}; (25) Esteban \etal \cite{esteban}. In the case of (8) and (16), we used the abundances derived from differential emission measure (DEM) reconstruction. 
}
\label{fig_ne_literature}
\end{figure}

Figure \ref{fig_ne_literature} shows the Ne abundances recently obtained for different objects in the solar vicinity (excluding peculiar and very evolved objects, such as Wolf-Rayet stars or planetary nebulae). Pre-main sequence stars were also omitted because the Ne abundances generally suffer from large uncertainties and the [Ne/O] ratios might be affected by depletion in the X-ray emitting material of oxygen onto dust grains (Drake \etal \cite{drake05}). This is an unbiased compilation in the sense that this is, to the best of our knowledge, a complete census of the values reported in the literature since Drake \& Testa (\cite{drake_testa}) study (we include, however, all B-star analyses discussed in Sect.~\ref{sect_comparison} and the value for the ionized gas in the Orion nebula derived by Esteban \etal \cite{esteban}). The oxygen abundance of AGS05 was adopted to obtain the absolute neon abundances from the coronal [Ne/O] ratios. For the stars analysed by Garc\'{\i}a-Alvarez \etal (\cite{garcia06}, \cite{garcia08}), Ness \& Jordan (\cite{ness_jordan}) and Wood \& Linsky (\cite{wood_linsky}), we used the photospheric values quoted in these papers. For the metal-poor stars \object{$\xi$ UMa} and \object{Ross 154}, we rescaled the solar abundance to the metallicities found by Cayrel de Strobel \etal (\cite{cayrel}; [Fe/H]$\sim$--0.35) and Eggen (\cite{eggen}; [Fe/H]$\sim$--0.24), respectively. Finally, in the case of the supermetallic binary system \object{$\alpha$ Cen} ([Fe/H]$\sim$+0.24), we used the oxygen abundance of Feltzing \& Gonzalez (\cite{feltzing_gonzalez}). We caution that the metallicities deviate significantly from solar in these three cases, and that the Ne abundances may not strictly apply to the Sun. A large scatter is observed in Fig.\ref{fig_ne_literature} (total spread of 0.42 dex), but it could be noted that the only few studies supporting a high solar Ne abundance are generally based on X-ray spectroscopy of strong coronal sources.

\section{Conclusion}\label{sect_conclusion}
A mean, absolute neon abundance, $\log \epsilon$(Ne)=7.97$\pm$0.07 dex, has been inferred from our combined NLTE abundance analysis of the photospheric \ion{Ne}{i} and \ion{Ne}{ii} lines in a sample of 18 nearby, early B-type stars. This indicates a value for the Sun $\sim$35\%  higher than the new recommended solar abundance (7.84$\pm$0.06 dex; AGS05). In contrast, an increase of the Ne abundance by a factor $\sim$3 in the solar interior is needed to restore the agreement between the solar models and the helioseismological data. Our results therefore clearly suggest that solving this problem by simply adjusting the solar Ne abundance would probably require an increase of this quantity well beyond the range of plausible values (note that an enhancement of the other metal abundances to within their uncertainties may also be necessary; Bahcall \etal 2005). We point out that this is a robust conclusion attained regardless of the ion or $T_{\rm eff}$ scale chosen (Fig.\ref{fig_teff}). 

Neon is produced during carbon burning in the final stages of the evolution of massive stars and one may naturally expect the Ne abundances of young, B-type stars to be higher than the solar value because of chemical enrichment over the past 4.6 Gyr. However, the predicted enhancements in the solar neighbourhood are likely to be very small according to Galactic chemical evolutionary models ($\sim$0.04 dex; Chiappini \etal \cite{chiappini}). Our mean Ne abundance should therefore be very close to the value prevailing in the protosolar nebula. Unexpectedly, however, metal abundances derived for nearby B stars are often found to be slightly (but significantly) below the most recent estimates for the Sun or the meteoritic values (e.g. Morel \cite{morel08}, and references therein). This discrepancy may be related to missing physics or unaccounted systematic errors in the B star analyses, and may question the assumption that the neon abundance derived from hot stars is directly transposable to the Sun. The fact that our mean Ne abundance is indistinguishable from the values recently determined for the ionized gas in the Orion nebula (8.05$\pm$0.07 dex; Esteban \etal \cite{esteban}), within supergranules (7.89$\pm$0.13 dex; Young \cite{young}) or from {\it in situ} observations of the solar wind (7.96$\pm$0.13 dex; Bochsler \cite{bochsler}) nevertheless suggests that the abundances we derived are representative of the solar value (see Fig.\ref{fig_ne_literature}). 

\begin{acknowledgements}
Thierry Morel acknowledges financial support from the Research Council of Leuven University through grant GOA/2003/04 and Belspo for contract PRODEX GAIA-DPAC. We are grateful to Nicolas Grevesse for interesting discussions about the neon abundance of the Sun and useful comments on the first draft of this paper. It is also a pleasure to thank Giusi Micela for precisions about coronal abundances. The careful reading of the manuscript and the useful suggestions from the anonymous referee were greatly appreciated. We are indebted to all the observers who contributed to the acquisition of the data used in this paper. The archival ELODIE data have been processed within the PLEINPOT environment.This research made use of NASA's Astrophysics Data System Bibliographic Services, the SIMBAD database operated at CDS, Strasbourg (France). 
\end{acknowledgements}

\appendix

\section{Examples of comparison between observed and synthetic Ne line profiles} \label{sect_line_profiles}
\begin{figure*}
\centering
\includegraphics[width=18.5cm]{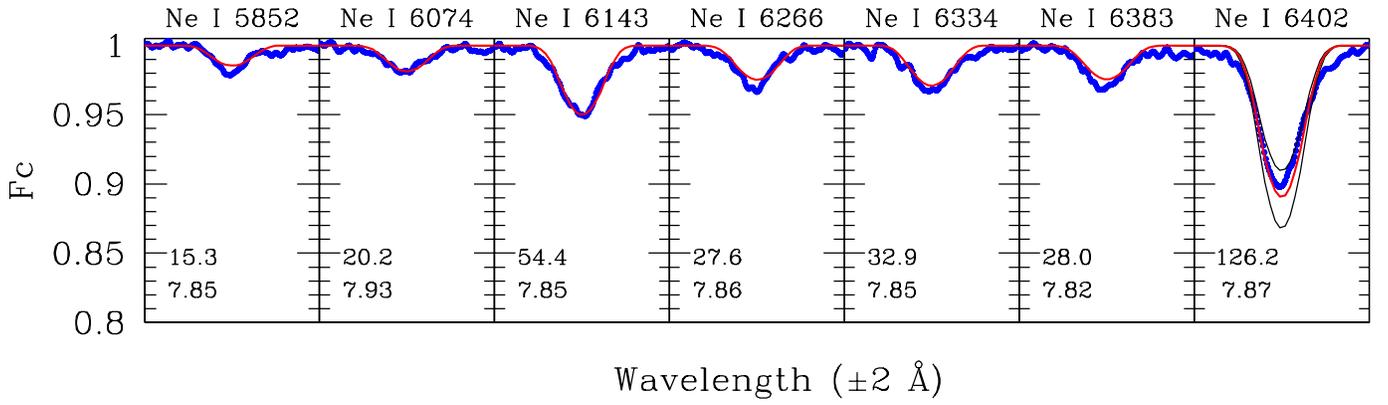}
\caption{Comparison for \icma \ between the observed ({\em blue dots}) and the synthetic profiles ({\em red line}) computed for the abundance given by the corresponding spectral line (value given in every panel, along with the measured EW; see Table~\ref{tab_ews}). The synthetic spectra were convolved with a rotational broadening function with $v \sin i$ = 32 km s$^{-1}$ (Morel \etal \cite{morel08}). The effect of varying the abundance by $\pm$0.1 dex is shown for \ion{Ne}{i} $\lambda$6402. The spectral range in each panel is 4 \AA.}
\label{fig_ne_profiles1}
\end{figure*}

\begin{figure*}
\centering
\includegraphics[width=18.5cm]{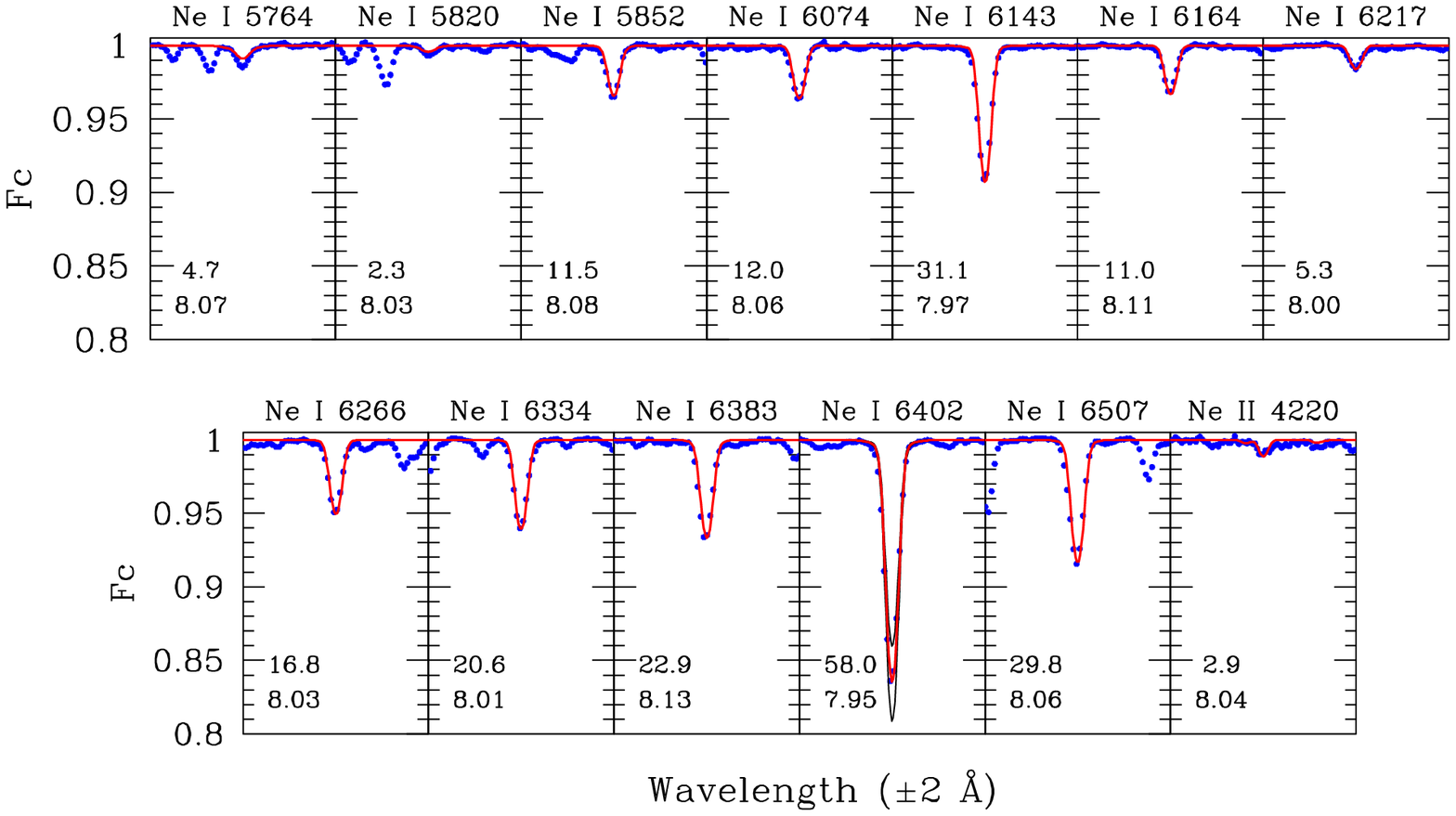}
\caption{Same as Fig.\ref{fig_ne_profiles1}, but for \gpeg. The synthetic spectra were convolved with a rotational broadening function with $v \sin i$ = 10 km s$^{-1}$ (Morel \etal \cite{morel08}).}
\label{fig_ne_profiles2}
\end{figure*}

\begin{figure*}
\centering
\includegraphics[width=18.5cm]{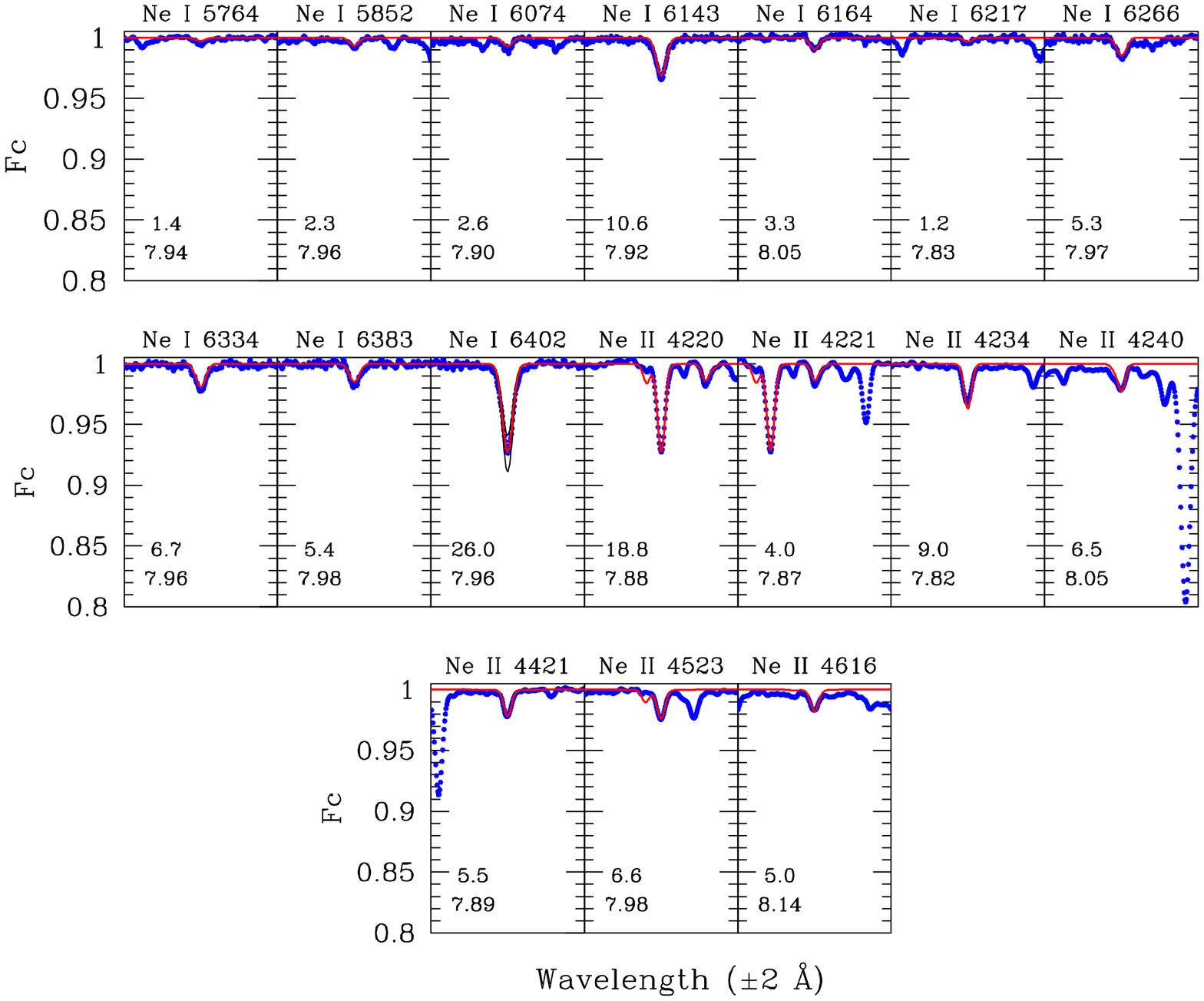}
\caption{Same as Fig.\ref{fig_ne_profiles1}, but for \xcma. The synthetic spectra were convolved with a rotational broadening function with $v \sin i$ = 10 km s$^{-1}$ (Morel \etal \cite{morel08}). \ion{Ne}{ii} $\lambda$4392 has been omitted in this figure, as it falls in the wing of \ion{He}{i} $\lambda$4387.9 (which is not included in the model calculations).}
\label{fig_ne_profiles3}
\end{figure*}

\begin{figure*}
\centering
\includegraphics[width=18.5cm]{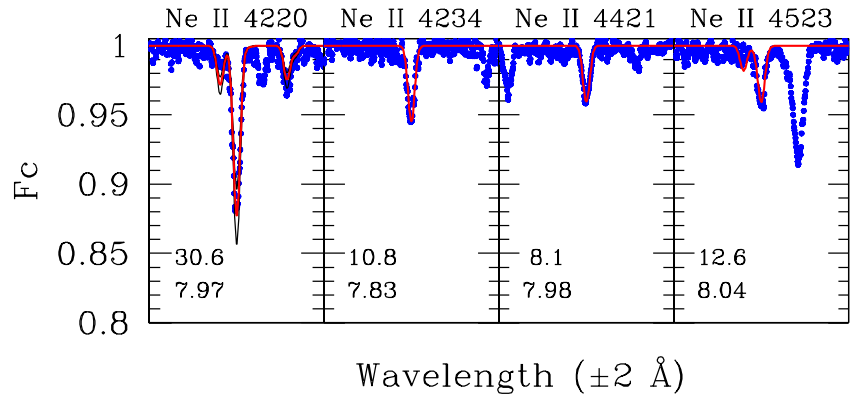}
\caption{Same as Fig.\ref{fig_ne_profiles1}, but for \tsco. The synthetic spectra were convolved with a rotational broadening function with $v \sin i$ = 8 km s$^{-1}$ (Hubrig \etal \cite{hubrig}). The effect of varying the abundance by $\pm$0.1 dex is shown for \ion{Ne}{ii} $\lambda$4220. \ion{Ne}{ii} $\lambda$4392 has been omitted in this figure, as it falls in the wing of \ion{He}{i} $\lambda$4387.9 (which is not included in the model calculations).}
\label{fig_ne_profiles4}
\end{figure*}

Figures~\ref{fig_ne_profiles1}--\ref{fig_ne_profiles4} show comparisons between the observed and the synthetic Ne line profiles for four representative stars sampling the whole $T_{\rm eff}$ range and with different data quality: \icma, \gpeg, \xcma \ and \tsco. 

\Online

\section{$\log gf$ values, EW measurements and line-by-line abundances} \label{sect_ews}
Table~\ref{tab_ews} lists the adopted $\log gf$ values, measured EWs and individual abundances for the neon lines used in the analysis. 

\onltab{1}{
\begin{table*}
\centering
\caption{Adopted $\log gf$ values, EW measurements (in m\AA) and line-by-line abundances (on the scale in which $\log \epsilon$[H]=12). }
\label{tab_ews}
\hspace*{-1.4cm}
\begin{tabular}{lrcccccccccccccccccc} \hline\hline
Transition$^a$ & $\log gf$ & \multicolumn{2}{c}{HD 886} & \multicolumn{2}{c}{HD 16582} & \multicolumn{2}{c}{HD 29248} &  \multicolumn{2}{c}{HD 30836} & \multicolumn{2}{c}{HD 35468} & \multicolumn{2}{c}{HD 36591} & \multicolumn{2}{c}{HD 44743} & \multicolumn{2}{c}{HD 46328} &  \multicolumn{2}{c}{HD 50707}\\
 & & \multicolumn{2}{c}{\gpeg} & \multicolumn{2}{c}{\dcet} & \multicolumn{2}{c}{\neri} & \multicolumn{2}{c}{\piori} & \multicolumn{2}{c}{\gori} & & & \multicolumn{2}{c}{\bcma} & \multicolumn{2}{c}{\xcma} & \multicolumn{2}{c}{\15cma}\\
\hline
{\bf \ion{Ne}{i}} & & & & & & & & & & & & & & & & & & &\\
5764.42      & --0.316 &  4.7 & 8.07 &  2.9 & 7.89 &  3.2 & 7.98        & ...  & ...  & ...  &  ...  &  ...  &  ...       &   2.0 & 7.83 &   1.4 & 7.94 &  ...  &  ...      \\
5820.16      & --0.517 &  2.3 & 8.03 & ...  & ...  & ...  & ...         & ...  & ...  &  ... & ...   & ...   &  ...       &  ...  &  ...       &  ...  &  ...       &   ... & ...       \\
5852.49      & --0.453 & 11.5 & 8.08 & 10.1 & 8.06 & ...  & ...         & 13.3 & 8.08 & ...  &  ...  & ...   &  ...       &  ...  &  ...       &   2.3 & 7.96 & ...   &    ...    \\
6074.34      & --0.508 & 12.0 & 8.06 & 10.0 & 8.01 & ...  & ...         & ...  & ...  & ...  & ...   &   4.7 &  7.96 &  ...  &   ...      &   2.6 & 7.90 &   ... &  ...      \\
6143.06      & --0.069 & 31.1 & 7.97 & 27.3 & 7.94 & 28.1 & 7.96 & 33.1 & 7.92 & 32.6 & 7.91 &  14.2 & 7.88  &  18.0 & 7.85 &  10.6 & 7.92 &  19.7 & 7.92\\
6163.59      & --0.613 & 11.0 & 8.11 &  9.2 & 8.06 &  8.2 & 8.04 & 10.6 & 8.01 & ...  &  ...         &  ...  &   ... &   5.0 & 7.92 &   3.3 & 8.05 &  ...  &    ...    \\
6217.28      & --0.936 &  5.3 & 8.00 &  4.2 & 7.93 & ...  & ...         & ...  & ...  & ...  &   ... &  ...  &   ...      &  ...  &  ...       &   1.2 & 7.83 & ...   &   ...     \\
6266.49      & --0.334 & 16.8 & 8.03 & 15.2 & 8.01 & 16.1 & 8.06 & 24.6 & 8.11 & 18.9 & 8.00 &   6.7 & 7.91  &  ...  &   ...      &   5.3 & 7.97 &  ...  &   ...     \\
6334.43      & --0.324 & 20.6 & 8.01 & 19.8 & 8.03 & 18.3 & 8.00 & 26.3 & 8.05 & 27.8 & 8.08 &  ...  &   ...      &  11.1 & 7.88 &   6.7 & 7.96 &  ...  &   ...     \\
6382.99      & --0.228 & 22.9 & 8.13 & 19.7 & 8.09 & 16.4 & 8.03 & 30.6 & 8.18 & 24.5 & 8.08 &   9.7 & 8.04  &   9.9 & 7.92 &   5.4 & 7.98 &  13.9 & 8.10\\
6402.25      &   0.336 & 58.0 & 7.95 & 55.3 & 7.96 & 57.1 & 7.92 & 72.7 & 7.93 & 65.0 & 7.85 &  31.5 & 7.89  &  41.7 & 7.86 &  26.0 & 7.96 &  38.3 & 7.87\\
6506.53      & --0.026 & 29.8 & 8.06 & 22.6 & 7.96 & 23.1 & 7.98 & ...  & ...  & ...  & ...   &  12.3 & 7.95 &  16.3 & 7.94 & ...   &   ...      &  ...  &   ...     \\
{\bf \ion{Ne}{ii}} & & & & & & & & & & & & & & & & & & &\\
4219.74      &   0.674 &  2.9 & 8.04        & ...  & ...         &  6.2 & 7.99        & ...  &  ...  & ...   & ...        &  14.1 & 7.94       &  10.2 & 7.84 &  18.8 & 7.88 &  16.3 & 8.05\\
4220.89      & --0.179 &  ... & ...         & ...  & ...         &  ... & ...         & ...  &  ...  & ...   & ...        &  ...  &  ...       &  ...  &   ...&   4.0 & 7.87 &  ...  &   ...     \\
4233.85      &   0.362 &  ... & ...         & ...  & ...         &  ... & ...         & ...  &  ...  & ...   & ...        &   7.0 & 7.95       &  ...  &   ...&   9.0 & 7.82 & ...   &  ...      \\
4240.10      & --0.089 &  ... & ...         & ...  & ...         & ...  & ...         & ...  &  ...  & ...   & ...        &  ...  &  ...       &  ...  &  ... &   6.5 & 8.05 &  ...  &  ...      \\
4391.99$^b$  &   0.908 &  ... & ...         & ...  & ...         & ...  & ...         & ...  &  ...  & ...   &  ...       &  ...  &  ...       &  13.4 & 7.93 &  25.5 & 7.93 &  12.2 & 7.87\\
4421.39      &   0.158 &  ... & ...         & ...  & ...         & ...  & ...         & ...  &  ...  & ...   & ...        & ...   &  ...       &   2.2 & 7.82 &   5.5 & 7.89 & ...   & ...       \\
4522.72      &   0.154 &  ... & ...         & ...  & ...         & ...  & ...         & ...  &  ...  & ...   & ...        & ...   &  ...       &   3.8 & 7.95 &   6.6 & 7.98 & ...   & ...       \\
4616.09      & --0.117 &  ... & ...         & ...  & ...         & ...  & ...         & ...  &  ...  & ...   & ...        & ...   & ...        &  ...  &  ... &   5.0 & 8.14 &  ...  &   ...     \\
\hline   
 & & & & & & & & & \\
\hline\hline
Transition$^a$ & $\log gf$ & \multicolumn{2}{c}{HD 51309} & \multicolumn{2}{c}{HD 52089} & \multicolumn{2}{c}{HD 129\,929} &  \multicolumn{2}{c}{HD 149\,438} & \multicolumn{2}{c}{HD 163\,472} & \multicolumn{2}{c}{HD 170\,580} & \multicolumn{2}{c}{HD 180\,642} & \multicolumn{2}{c}{HD 205\,021} & \multicolumn{2}{c}{HD 214\,993}\\
 & & \multicolumn{2}{c}{\icma} & \multicolumn{2}{c}{\ecma} & \multicolumn{2}{c}{\vcen} & \multicolumn{2}{c}{\tsco} & \multicolumn{2}{c}{\voph} & & & & & \multicolumn{2}{c}{\bcep} & \multicolumn{2}{c}{\twelvelac}\\
\hline
{\bf \ion{Ne}{i}} & & & & & & & & & & & & & & & & & & &\\
5764.42      & --0.316 & ...  & ...        &  ... &    ...     &  ... &   ...      &  ... &    ...     &   ... &   ...      &   7.7 & 8.19 &  ...  &    ...     &  ...  &    ...     &   ...&     ...   \\
5820.16      & --0.517 & ...  & ...        & ...  &    ...     &  ... &    ...     &  ... &   ...      &  ...  &    ...     &  ...  &   ...      &  ...  &   ...      &   ... &  ...       &  ... &  ...      \\
5852.49      & --0.453 & 15.3 & 7.85 & ...     &  ...       &  ... &   ...      & ...  &   ...      &  ...  &  ...       &  19.7 & 8.11 &  ...  &  ...       & ...   & ...        &  ...  &  ...       \\
6074.34      & --0.508 & 20.2 & 7.93 & ...     &  ...        &  ...    &  ...          &  ...    &    ...        &  ...     &   ...         &  23.2 & 8.15 &  ...     &  ...          &  ...     &   ...         & 11.0 & 8.14\\
6143.06      & --0.069 & 54.4 & 7.85 & 21.6 & 7.83 & 23.1 & 7.94 &   ...   &      ...      &  34.1 & 8.05 &  44.3 & 7.97 &  23.6 & 7.99 &  14.8 & 7.86 &   ...   &   ...        \\
6163.59      & --0.613 & ...  &  ...          &  ...    &    ...        & ...     &     ...       &  ...    &   ...         &  ...     & ...            &  19.2 & 8.18 & ...       & ...            & ...       & ...            &  7.7 & 8.06\\
6217.28      & --0.936 & ...  &  ...          &  ...    &   ...         &  ...    &   ...         &  ...    &   ...         &  ...     &  ...          &  11.3 & 8.14 &   ...    &    ...        & ...      &    ...        & ...     &       ...    \\
6266.49      & --0.334 & 27.6 & 7.86 & 14.0 & 7.97 & 14.6 & 8.09 &  ...    & ...           &  ...     & ...           &  23.2 & 7.99 &  ...     &  ...          &   ...    & ...           & 10.9 & 7.93\\
6334.43      & --0.324 & 32.9 & 7.85 & 12.6 & 7.83 & 21.0 & 8.14 &  ...    & ...           & ...      & ...           &  33.2 & 8.05 & ...      & ...           &  12.6 & 8.03 & 15.0 & 7.96\\
6382.99      & --0.228 & 28.0 & 7.82 & 11.7 & 7.88 & 14.1 & 8.03 &  ...    & ...           & ...      & ...           &  32.9 & 8.09 & ...      & ...           &  10.6 & 8.05 &   ...   &     ...      \\
6402.25      &   0.336 &126.2 & 7.87 & 48.6 & 7.82 & 42.5 & 7.86 &  ...    & ...           &  65.6 & 8.04 &  73.8 & 7.89 &  41.6 & 7.88 &  30.9 & 7.83 & 50.6 & 7.90\\
6506.53      & --0.026 & ...  &  ...          & 16.8 & 7.84 & 21.0 & 8.01 &  ...    &  ...          & ...      & ...           &  44.5 & 8.05 &   9.9 & 7.74 &  11.4 & 7.88 & 17.6 & 7.91\\
{\bf \ion{Ne}{ii}} & & & & & & & & & & & & & & & & & & &\\
4219.74      &   0.674 & ...  &  ...          &  7.2 & 7.86 &   ...   &   ...         & 30.6 & 7.97 &  ...     &      ...      &  ...     &   ...         &  16.5 & 8.02 &  15.3 & 7.95 &  9.4 & 7.98\\
4220.89      & --0.179 & ...  &  ...          & ...     &  ...          & ...     & ...           &  ...    &       ...     &    ...   &      ...      &   ...    &     ...       &  ...     &   ...         &   ...    &    ...        &  ...    &    ...       \\
4233.85      &   0.362 & ...  & ...           & ...     &  ...          & ...     & ...           & 10.8 & 7.83 &   ...    &     ...       &   ...    &     ...       &   5.6 & 7.87 & ...      &    ...        &  ...    &      ...     \\
4240.10      & --0.089 & ...  & ...           & ...     &  ...          & ...     & ...           &  ...    &      ...      &    ...   &      ...      &    ...   &        ...    &    ...   &     ...       & ...      &   ...         &  ...    &    ...       \\
4391.99$^b$  &   0.908 & ...  & ...           &  8.0 & 7.86 &  7.9 & 8.01 & 35.3 & 8.05 &  ...     &  ...          &   ...    &   ...         &  10.0 & 7.72 &  14.5 & 7.91 & ...     &   ...        \\
4421.39      &   0.158 & ...  & ...           & ...     &  ...          & ...     & ...           &  8.1 & 7.98 &  ...     &   ...         &  ...     &    ...        &   ...    &   ...         &    ...   &      ...      &  ...    &   ...        \\
4522.72      &   0.154 & ...  & ...           & ...     &  ...          & ...     & ...           & 12.6 & 8.04 &  ...     &   ...         &  ...     &   ...         &   ...    &   ...         & ...      &     ...       &   ...   &         ...  \\
4616.09      & --0.117 & ...  & ...           & ...     &  ...          & ...     &   ...         &  ...    &     ...       &  ...     &    ...        &  ...     &   ...         &  ...     &     ...       &  ...     &   ...         &  ...    &   ...        \\
\hline
\end{tabular}
\begin{flushleft}
Notes: A blank indicates that the EW was not reliably measurable, the line was considered blended for the relevant temperature range or yielded a discrepant abundance. The accuracy of the EW measurements is discussed in Sect.~\ref{sect_analysis}.\\ 
$^a$ The mean wavelength is given.\\
$^b$ Wing of \ion{He}{i} $\lambda$4387.9 taken as pseudo continuum.\\
\end{flushleft}
\end{table*}
}

\end{document}